\def\la{\mathrel{\mathpalette\fun <}}
\def\ga{\mathrel{\mathpalette\fun >}}

\def\fun#1#2{\lower0.837ex\vbox{\baselineskip0ex\lineskip0.209ex
  \ialign{$\mathsurround=0ex#1\hfil##\hfil$\crcr#2\crcr\sim\crcr}}}

\def\sles{\lower2pt\hbox{$\buildrel {\scriptstyle <}
   \over {\scriptstyle\sim}$}}

\def\sgreat{\lower2pt\hbox{$\buildrel {\scriptstyle >}
   \over {\scriptstyle\sim}$}}

\def\la{\mathrel{\mathpalette\fun <}}
\def\ga{\mathrel{\mathpalette\fun >}}

\documentclass{svjour3}                     
\smartqed  
\usepackage{graphicx}
\begin{document}

\title{Gamma-Ray Telescopes
}
\subtitle{400 Years of Astronomical Telescopes}


\author{Neil Gehrels  \and
      John K. Cannizzo
}


\institute{N. Gehrels  \at
              NASA/GSFC/ASD/Code 661 \\
              Greenbelt, Md 20071 USA \\
              Tel.: (301)286-6546\\
              \email{gehrels@milkyway.gsfc.nasa.gov}
           \and
         J. K. Cannizzo  \at
         CRESST/UMBC/NASA/GSFC/ASD/Code 661 \\
         Greenbelt, Md 20071 USA \\
}

\date{Received: date / Accepted: date}

\maketitle

\begin{abstract}
The last half-century has seen 
    dramatic developments
    in  $\gamma-$ray telescopes, 
   from their initial conception and development
       through  to their blossoming 
    into full maturity
   as a potent research tool in astronomy.
    Gamma-ray telescopes are leading research
 in diverse areas such as $\gamma-$ray bursts,
   blazars, Galactic transients,
    and
  the Galactic distribution of $^{26}$Al.
\end{abstract}

\section{Introduction}
\label{intro}
The current panoply of high-energy astronomical observatories
                          is impressive:
  {\it INTEGRAL}, {\it Swift}, {\it Fermi-GLAST},
   {\it RHESSI}, {\it Suzaku}, and {\it AGILE}.
  These satellites 
   have made fundamental contributions to their fields
  and in fact have had broad impact beyond their immediate
   purviews.
     These areas of study include $\gamma-$ray bursts
({\it Swift}, {\it INTEGRAL}, {\it Fermi-GLAST}), 
    solar physics ({\it RHESSI}),
  supernovae ({\it Swift}),
  Galactic astronomy ({\it INTEGRAL}, {\it Fermi-GLAST}),
   Galactic transients ({\it Swift}, {\it Suzaku}), 
  and AGN ({\it Suzaku}).
\section{Background}
\label{sec:1}
A basic consideration of the theory of photon
interaction as a function of energy divides
 naturally into three regimes:
  photoelectric interactions at $E \la 1$ MeV,
 Compton scattering for  $E \simeq 1$ MeV, and
pair production for $E \ga 1$ MeV
  (Figure 1).
\begin{figure}
\includegraphics[height=110mm]{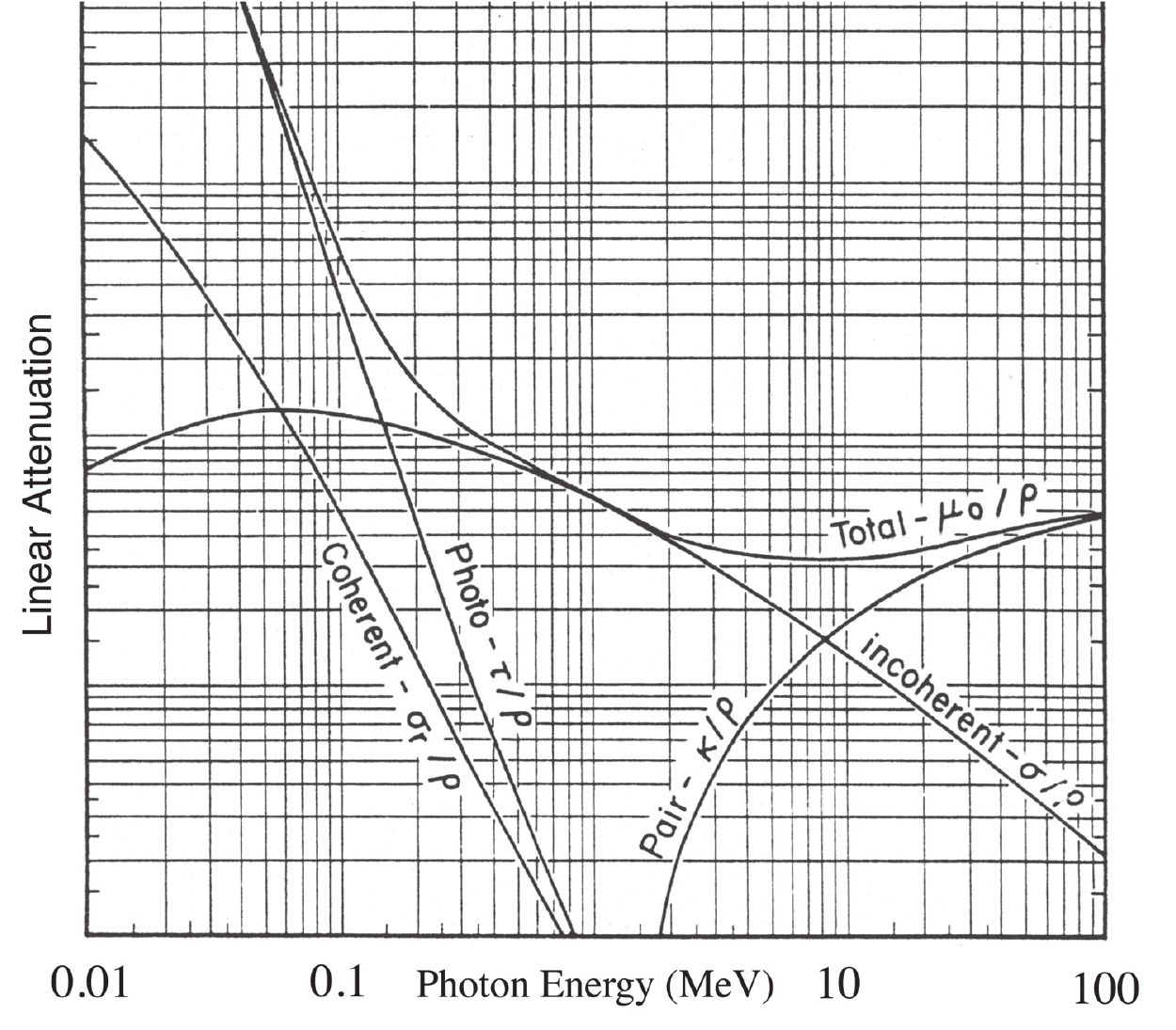}
\caption{A schematic showing of photon interactions
     indicating linear attenuation in an absorptive
   medium versus energy.}
\label{fig:1}   
\end{figure}
   This natural ordering has provided the impetus
  for three basic telescope designs, 
    (i) coded aperture mask for low energies, in which 
   the illumination of a non-redundant array grid casts a shadow
on a detector, thereby giving a sky position
   (e.g., the BAT on {\it Swift}),
   (ii)  a Compton telescope for  medium energies (e.g., COMPTEL on {\it CGRO},
   and (iii)  pair telescope in which an incident high-energy $\gamma-$ray
   interacts with a medium and produces a $e^+-e^-$ pair.
   The secondary  phase consists of signal detection,
  which  for the three cases 
 is carried out by 
  (i) scintillators (e.g., NaI, BGO, CsI, LaBr$_3$,
  and GSO),
   (ii)  solid state detectors
   (e.g., Si, CdZnTe, Ge, HgI$_2$),
  and (iii) pair trackers
   (e.g., spark chamber, Si).
   Figure 2 shows the history of $\gamma-$ray technology, with examples
   of observatories based
on these 3 strategies.
\begin{figure}
\includegraphics[height=81.5mm]{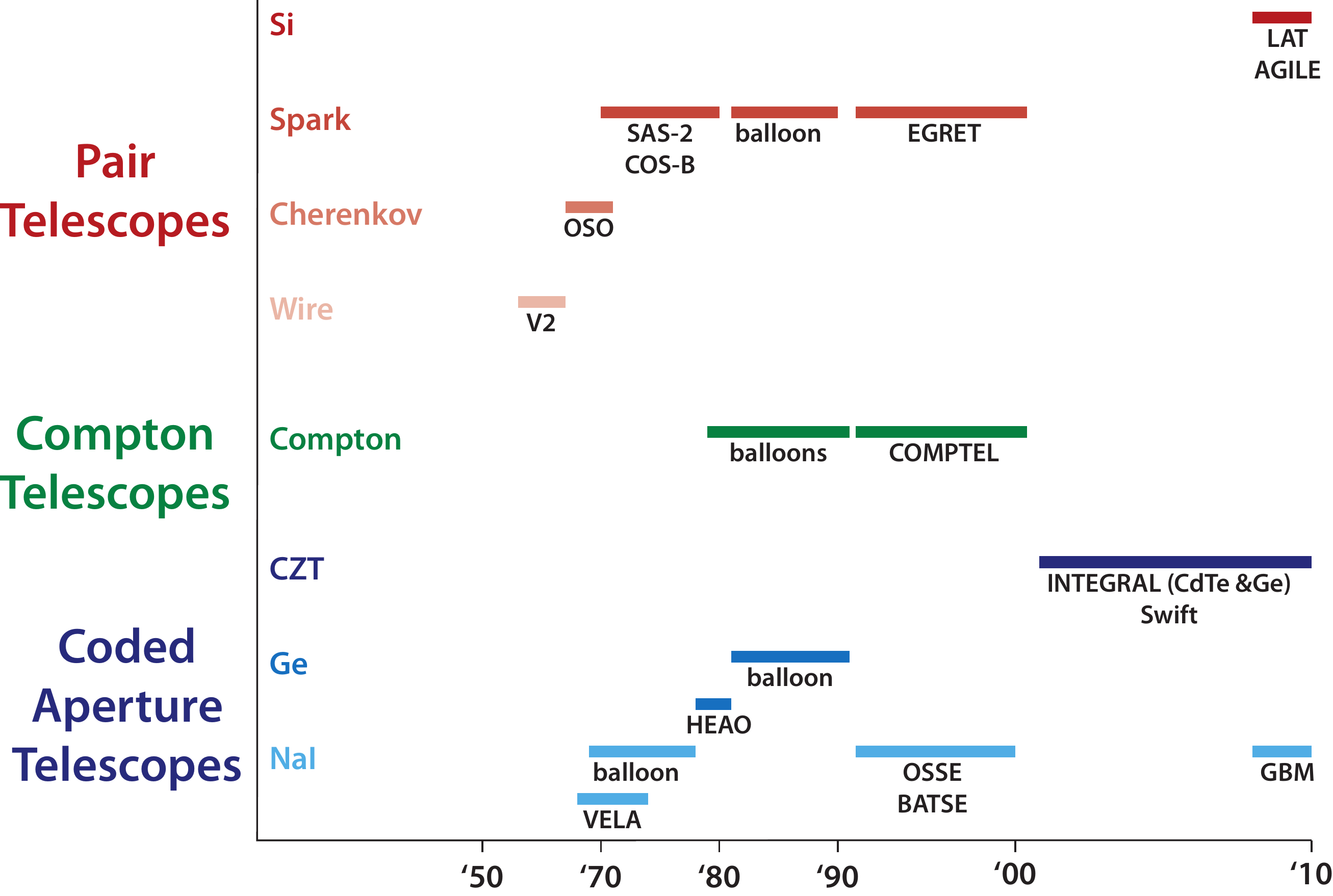}
\caption{A history of $\gamma-$ray telescope technology.}
\label{fig:2}   
\end{figure}

\section{Telescope Technologies}


\subsection{Low Energy Gamma Ray Telescopes}

The earliest astronomical spacecraft
   (actually -  unintentionally astronomical)
 to utilize collimated detectors 
were the {\it Vela} satellites
 (Klebesadel, Strong, \& Olson 1973).
Each carried six 10 cm$^3$ CsI scintillation counters 
distributed so as to achieve optimal nearly isotropic sensitivity.
The detectors were sensitive in the energy range between $0.2$ and $1.0$ MeV
for {\it Vela 5} and  $0.3-1.5$ MeV for {\it Vela} 6,
with a detector  efficiency  $0.17-0.50$.
  The scintillators were shielded
against direct penetration by electrons
  below $\sim0.75$ MeV
and protons below $\sim20$ MeV.
A high $Z-$shield
(i.e., a shield to protect against nuclei with large atomic number)
attenuated photons with energies 
  below that
of the counting threshold.
   No active anti-coincidence
 \footnote{In anticoincidence 
  counting, two counters 
are connected so that a pulse is recorded 
by one of them only if there is no simultaneous 
pulse in the other. This is useful in rejecting 
particles that do not originate 
from the source being studied.}
shielding was provided.
%
%
The accumulated data   included a background
component due to cosmic particles and
their secondary effects.
The observed background rate
was $\sim150$  c s$^{-1}$ for  {\it Vela} 5 
and  $\sim20$  c s$^{-1}$ for {\it Vela} 6.

{\it HEAO-I} (1977-1979)
    had gas-proportional counters,
with each counter possessing a multigrid
collimator (Wood et al 1984).
  Each collimator was composed of a stack of etched
Mo         sheets interleaved with spacer frames.
  To maintain thermal stability
  a heat shield made of 2 $\mu$m Kimfol
   polycarbonate film and coated
 with 80 nm of Al 
    was put
   in front of each collimator.
   Incident X-rays passed through the heat shield
and Mylar film, with the net
transmission of the two layers determining the response
to soft X-rays.

Collimation of $\gamma-$rays
  and background
reduction 
for the OSSE instrument
(Johnson et al. 1993)
 aboard {\it CGRO}   (1991-2000)
were achieved via active
shielding and a passive
W      collimator.
Each of the four OSSE detectors
  consisted of a  NaI scintillation
 crystal
  shielded behind
 by an optically coupled
7.6 cm thick CsI scintillation
 crystal in a ``phoswich'' (phosphor sandwich) configuration.
Each phoswich was enclosed
  in an annular shield of NaI
scintillation
 crystals providing anticoincidence
  for $\gamma-$ray interactions in the phoswich.
The annular shield also enclosed a W
 slat collimator which sets the $3.8^{\circ} \times 11.4^{\circ}$
FWHM $\gamma-$ray FOV of the phoswich detector.

The BAT instrument on {\it Swift}  (2004-)
 has a D-shaped coded mask\footnote{
The concept of the coded mask was 
discovered by Aristotle. As he was 
walking under some trees, he noticed 
that patches of sunlight on the 
ground were in the shape of the 
Sun and not of the gap between 
the branches. The trees acted 
in the same way as a coded mask.}
 made of $\sim54,000$ Pb
 tiles ($5\times5\times 1$ mm)
mounted on a 5 cm thick composite honeycomb
panel 
 which is mounted by composite fiber struts 1 m
above the detector plane
(Gehrels et al. 2004).
 The BAT coded mask uses a random 50\% open/50\% closed
  pattern rather than the more common  uniformly redundant array pattern.
The mask area is 2.7 m$^2$, which yields a half-coded
FOV of 1.4 sr.

\subsection{Pair Telescopes}


  OSO-3 (1967-1968)
   achieved
  the first extensive 
   $\gamma-$ray observation in space,
 using a Cerenkov scintillator detector 
  (Badhwar, Kaplon, \& Valentine 1974),
    and
    showed  the Galactic disk to be  a strong
   source of high-energy $\gamma-$rays.
 The detector  consisted of 0.64 cm thick plastic scintillators
and a 1 in thick lucite  Cerenkov radiator
  over a 0.16   cm thick Pb sheet to convert
$\gamma-$rays into $e^{-}-e^{+}$ pairs.
  The discriminator thresholds on all detectors were
set using sea level muons.
 The energy threshold was 50 MeV (1\%), rising
to 13.5\% at 200 MeV, and 17\% at 1 GeV,
   where the percentages indicate the fraction 
 of the total geometric area.
 The false alarm rate for singly charged particles
was about  0.15\%.


The $\gamma-$ray telescope flown on  SAS-2 (1972-1973)
  consisted of an assembly with 16 spark
  chamber modules above a set of four plastic scintillators,
with another 16 modules below the scintillators
  (Fichtel, Kniffen, \& Hartman 1973).
  Thin W plates 
 (which had a thickness of  $\sim0.03$ radiation lengths)
were interleaved between the spark chamber modules,
  with an area $\sim640$ cm$^2$.
  The large number of W plates and spark chambers
provided material for an incoming $\gamma-$ray to be converted
into a $e^{-}-e^{+}$ pair, which could  then be identified
   and used to provide a direction for the incoming $\gamma-$ray,
and a way of determining the energy of the electrons
in the pair  by measuring the Coulomb scattering.
  The sensitivity range of the instrument was about $30 - 200$ MeV.


 After {\it Vela}, 
  COS-B (1975-1982) was the primary workhorse for studying $\gamma-$ray
bursts   (Boella et al. 1976).
  COS-B consisted of a dome-shaped
  1.1m$^2 \times 10$mm
  thick plastic scintillator
   viewed by nine photomultipliers for rejection of charged 
particles (the anti-coincidence counter, or ACO).
 The ACO maximum sensitivity  was at 700 KeV.


The EGRET instrument (Kanbach et al. 1988)
  on {\it CGRO} 
  consisted of an anticoincidence system,
  a spark chamber
  with
   interspersed  conversion material
  to materialize the incident photons and
  determine the trajectory of the secondary $e^{-}-e^{+}$ pair,
  a triggering device to detect the presence  
  of charged particles with the correct direction,
   and an energy  measuring device - 
   a total-absorption spectrometer NaI crystal.
  The 2-cm-thick plastic scintillators
  were viewed by 24 photomultiplier tubes
  coupled to the lower perimeter
   so as to keep the EGRET FOV
   free of irregular mass distributions.
   The conversion of the incoming $\gamma-$rays
  into $e^{-}-e^{+}$ pairs occurred
   in an upper stack of 28 spark chambers
   interleaved with Ta foils with an average thickness
  of 90 $\mu$m ($=0.02$ radiation lengths).
    The total depth for on-axis radiation was hence
  about 0.54  radiation lengths, leading to
  a conversion efficiency of about 30\% at 100 MeV.
   The volume of the upper chamber was $\sim80\times80\times45$
  cm$^3$, so that
   an effective area $\sim2000$ cm$^2$
  resulted for the center of the FOV.


  The LAT (Atwood et al. 2009)
    on board {\it Fermi-GLAST} (2008-)
 is a pair conversion telescope
  with a precision converter-tracker
  and calorimeter, each consisting
  of a $4\times4$ array of 16 modules
   supported by a low-mass Al grid structure.
  A segmented anticoincidence detector
  covers the tracker array, and a programmable
trigger and data acquisition system
  uses prompt signals from the tracker,
  calorimeter, and anticoincidence detector
  subsystems to form a trigger.
  The precision converter-tracker
   has 16 planes of high-Z material
  in which $\gamma-$rays incident on the LAT
  can convert to an $e^{-}-e^{+}$ pair.
  The converter planes are interleaved with
  position-sensitive detectors that record
  the passage of charged particles,
  thus measuring the tracks resulting from 
  $e^{-}-e^{+}$ pair conversion.
   This information is used to reconstruct
  the directions 
  of the incident $\gamma-$rays,
   Each tracker module has 18 $x$, $y$ tracking
  planes, consisting of two  layers ($x$ and $y$)
  of single-sided Si strip detectors.
   The 16 planes at the top of the tracker
    are interleaved with high-Z convert material (W),
  i.e., material capable of generating interactions 
             with nuclei of high atomic number.

\subsection{Compton Telescopes}

The COMPTEL instrument (Sch\"onfelder 1991)
  on {\it CGRO} 
  combined a large  FOV
  with imaging.
  It consisted  of two detector arrays,
and upper one of low-Z material
 and a lower one of high-Z material.
   In the upper detector an incoming
  $\gamma-$ray was  Compton scattered,
  and the scattered photon subsequently made
a second interaction in the lower detector.
 A time-of-flight measurement was utilized to
confirm the causal relation.
  The locations and energy losses of both 
interactions were monitored.
   For events that were completely absorbed,
the arrival direction of the $\gamma-$ray
  had to lie within a known angle
of the direction of the scattered $\gamma-$ray,
  which allowed  a determination of its original direction on the sky.
  Incompletely absorbed events yielded uncertainty cones
  that did  not constrain the source position.

\subsection{Focusing Telescopes}


Actual imaging of hard X-rays
 ($5-80$ keV)
  can be achieved
  using   
  multilayer mirrors.
   The technology
 has been proven
  with the InFOC$\mu$S balloon instrument,
  which obtained a 2.6 arcmin 
  resolution image of 4U 0115
(Tueller et al 2005).
   The Pt/C reflector surface
   multilayer mirror
   had a diameter of 40 cm and a focal
length of 8 m.
   The incident angle could  lie 
 in the range $0.105-0.356$ deg.
     There were 255 nestings and 2040 reflectors.
  The overall telescope had an effective
  area of 49 cm$^2$ at 30 keV,
   an angular resolution of 2.4 arcmin (half-power
diameter), and a field of view of 
  11 arcmin.

Currently under development is the
{\it NuSTAR} instrument, a NASA SMEX mission slated for
launch in $\sim2011$
which will contain an extensible boom for its
multilayer mirrors, and CdZnTe detectors
(Harrison et al. 2005).
  {\it HEFT},  a proof-of-concept 
  balloon payload,
    contains three
co-aligned conical-approximation
Wolter I mirrors,
each of which focuses hard
X-rays/soft $\gamma-$rays onto
a shielded solid-state
CsZnTe focal plane.
Each of the three optics
has 70 shells,
with radii between 4 and 12 cm.
The reflectors are coated with 
W/Si multilayers,
providing good reflectance
up to the W K-edge
at 69.5 keV.
The total collecting area
on-axis at 30 keV is 100 cm$^2$.
 {\it NuSTAR} consists of three co-aligned
telescope modules with a 10-m focal length,
and the optics and detectors are placed at either 
end of an extensible  mast.
The optics
   consist of 130 shells each, with radii between
5.5 and 16.9 cm.
 The shells are coated with a combination of W/SiC and Pt/SiC
multilayers.
   Combined with low graze angles,
the smooth response goes up to 80 keV.
 {\it  NuSTAR} will achieve angular resolution
of 40 arcsec (half-power diameter).
   The  {\it NuSTAR} and {\it HEFT}
   focal plane detectors have the same dimensions,
but the {\it NuSTAR} detectors are surrounded by an active
CsI shield rather than graded-Z/plastic shield.

 
   Looking further into the future, 
             new methods are
being discussed that could in principle take advantage
of the very short wavelength  of $\gamma-$rays to
achieve ultra-high resolution imaging.
The diffraction-limited angular resolution
\begin{equation}
\theta_d = 1.22 {\lambda\over d} = 0.12 \left( {500 \ {\rm keV}}\over E \right) \left({5 \ {\rm m}}\over d\right)
  \ \mu{\rm arcsec},
\end{equation}
  where $d$ is the lens diameter.
    This ultra-high resolution could be applied to directly
  image the event horizons of black holes.
  One technique for accomplishing direct, high-resolution imaging
is through Fresnel lenses (Skinner et al. 2004).
Fresnel zone plates (FZPs)
  focus radiation by blocking with opaque annuli
  the photons contributing to 
spatial frequencies that would arrive at the image plane
    non-optimally.  This 
technique is similar in spirit to ``apodization''
(literally - ``reducing the feet'' - referring to the diffraction pattern)
used by optical astronomers in which a PSF is enhanced 
by minimizing the higher order maxima of the diffraction pattern,
   effectively by throwing more of the light into the central,
   on-axis maximum.
 Phase Zone Plates (PZPs) go one step better by replacing the opaque
annuli with media possessing a thickness of refractive material
 which imposes a phase shift of $\pi$.
   The final step in improvement consists of  Phase Fresnel Lenses (PFLs)
in which the refractive material has a profile that is radial with respect 
to the central optical axis, so that all the radiation arrives at
the focal point with exactly the correct phase.
   This is an exact analog of the optical astronomers' apodization.
 The use of PFLs for high-energy photons
  is achievable because all materials have a refractive
  index that differs from unity.
  For X-rays and $\gamma-$rays the refractive index  $\mu=(1-\delta)$,
where $\delta$ is a small positive number, leading to a $\mu$
value slightly less than unity.

Another possibility 
for direct imaging is through
the use of
     Laue concentrators
  (von Ballmoos et al. 2004).
 This method takes advantage
of the phase information carried
  within the high-energy photons
   by using Bragg diffraction,
which refers to the interference
between the periodic (wave) nature of light
 and the  periodic structure
  of matter in a crystal lattice.
   In a Laue diffraction lens, a large
number of crystals  are oriented 
  in an annular configuration
  so that
incident photons  are imaged onto a common
focal spot.
   The background noise is kept low because
 of the resultant small detector volume.
  In a Laue-type crystal diffraction lens
the crystals are positioned along concentric rings.
  In order to be diffracted the wavelength
and incident angle of the incoming photon
must obey the Bragg relation
\begin{equation}
2d\sin\theta_B=n\lambda,
\end{equation}
where $d$  is the crystal plane spacing, $\theta_B$ the
incident angle of the photon, $n$ the reflection order,
and $\lambda$ the photon wavelength.
  A crystal at a distance $r_1$ from the optical axis
  is oriented such that the angle between the incident
beam and the crystalline planes
 is the Bragg angle $\theta_B$.
   For a given focal distance $f$ of the lens,
the radius of the $i$th ring is given by
\begin{equation}
r_i=f\tan|2\theta_{B_{i}}| \approx f {n\lambda\over d_i},
\end{equation}
   where $d_i$ is the crystalline plane spacing of the $i$th ring.
  The bandwidth
\begin{equation}
\Delta E \approx {2 E^2 d \Delta\theta\over nhc}
          \approx 40 \ {\rm keV}
        \left( E \over 511 \ {\rm keV} \right)^2
        \left( d\over d_{{\rm Ge[111]}} \right)
        \left( \Delta\theta \over 1^{'} \right),
\end{equation}
  where $\Delta\theta$ is the angular range over
   which the crystals reflect monochromatic radiation,
  and  Ge[111] refers to the
   germanium $[hkl]=[111]$ crystalline plane.

\section{History}

\begin{figure}
\includegraphics[height=70mm]{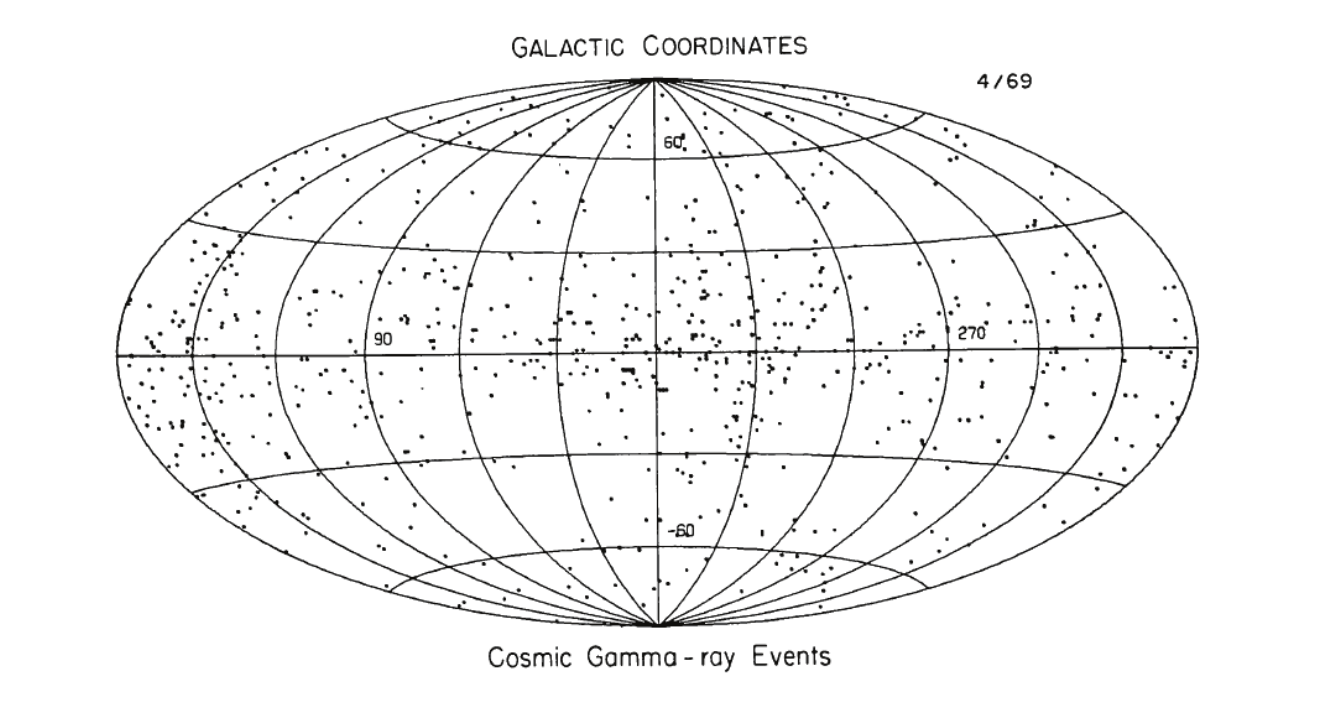}
\caption{The first $\gamma-$ray map, from Kraushaar et al. 1968.}
\label{fig:3}  
\end{figure}
The first $\gamma-$ray detector was launched on a V2 rocket as a
suborbital flight.
  It was sensitive in the range $3.4-90$ MeV
 and used wire ionization detectors.
  The astrophysical findings were upper limits
  (Perlow \& Kissinger 1951ab).
  The first $\gamma-$ray maps
 were made by {\it Explorer-11} and {\it OSO-3}.
   The high-energy
   instruments on these satellites
     were of the cathode-ray type, and their
  design foreshadowed future directions in their
  use of converter, calorimeter, and plastic 
  anticoincidence (Kraushaar, Clark, \& Garmire  1968).
 Figure 3
  shows the
first $\gamma-$ray map, 
   a  distribution  
  characterized by 
  a broad maximum toward the Galactic
   center, and an isotropic
  component with a softer
   energy spectrum.
   The spark-chamber era, spanning 20 MeV $-$ 10 GeV,
  was dominated by COS-B from 1975 to 1981,
 and by EGRET on {\it CGRO} from 1991 to 2000.
   The current instrument continuing this thread is
 the LAT on {\it Fermi-GLAST},
  with $\sim 10^6$ channels and 70 m$^2$ of Si surface.
    The Galactic map obtained by EGRET
  has yielded a plethora   
   of Galactic sources  and enriched our knowledge
  of their high-energy behavior.
The third EGRET catalog (Figure 4),
  which contains
data from 1991 April 22 to 1995 October 3,
has 271 sources ($E>100$ MeV).
  The list includes
  a 
  solar flare,
the LMC, five pulsars, 
one probable radio galaxy detection (Cen A),
66 blazars,
and  170 unidentified EGRET 
  sources   (Hartman et al. 1999).
Model predictions indicate
that {\it Fermi-}LAT should be 
able to detect 9900 sources at $>5\sigma$ 
in a one-year all-sky survey.
\begin{figure}
\includegraphics[height=155mm]{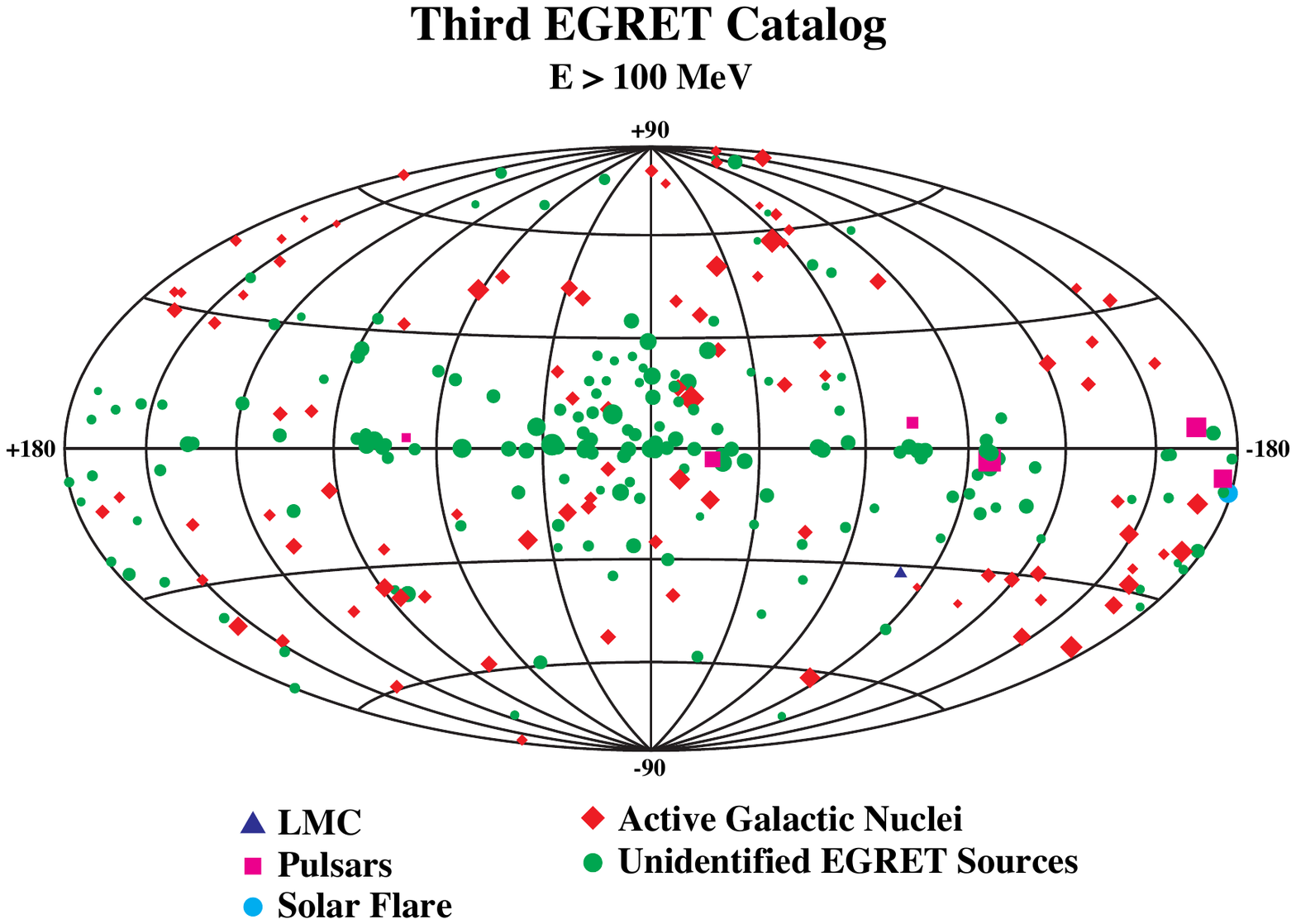}
\caption{The third EGRET catalog (Hartman et al. 1999).}
\label{fig:4}  
\end{figure}

Early Compton telescopes ($1-30$ MeV)
 were flown on balloon programs at MPE in Germany, 
and UC Riverside and UNH in the US.
  This unexplored energy range
  is technically challenging due to scattering,
 and localizations on the sky are typically to $\sim4^{\circ}$.
A major advance was the COMPTEL instrument on {\it CGRO} 
which operated from 1991 to 2000 
(Sch\"onfelder et al. 1993).
 A new window to astronomy
  was hereby opened:
    COMPTEL produced
  a reliable      $1-30$ MeV sky map, 
   shown in Figure 5.
(Sch\"onfelder et al. 2000).
This was the first complete $0.75-30$ MeV all-sky
  survey,
  which found
    32 non-variable sources and 31 $\gamma-$ray bursters. 
  Among the continuum sources were           pulsars 
                that had spun down,
 stellar black-hole candidates, 
supernova remnants, interstellar clouds, AGN, 
    $\gamma-$ray bursters, and the Sun during solar flares.
  Line detections were also made: 
          1.809 MeV ($^{26}$Al), 
          1.157 MeV ($^{44}$Ti), 
0.847 and 1.238 MeV ($^{56}$Co),  
          2.223 MeV (neutron capture). 
\begin{figure}
\includegraphics[height=58.5mm]{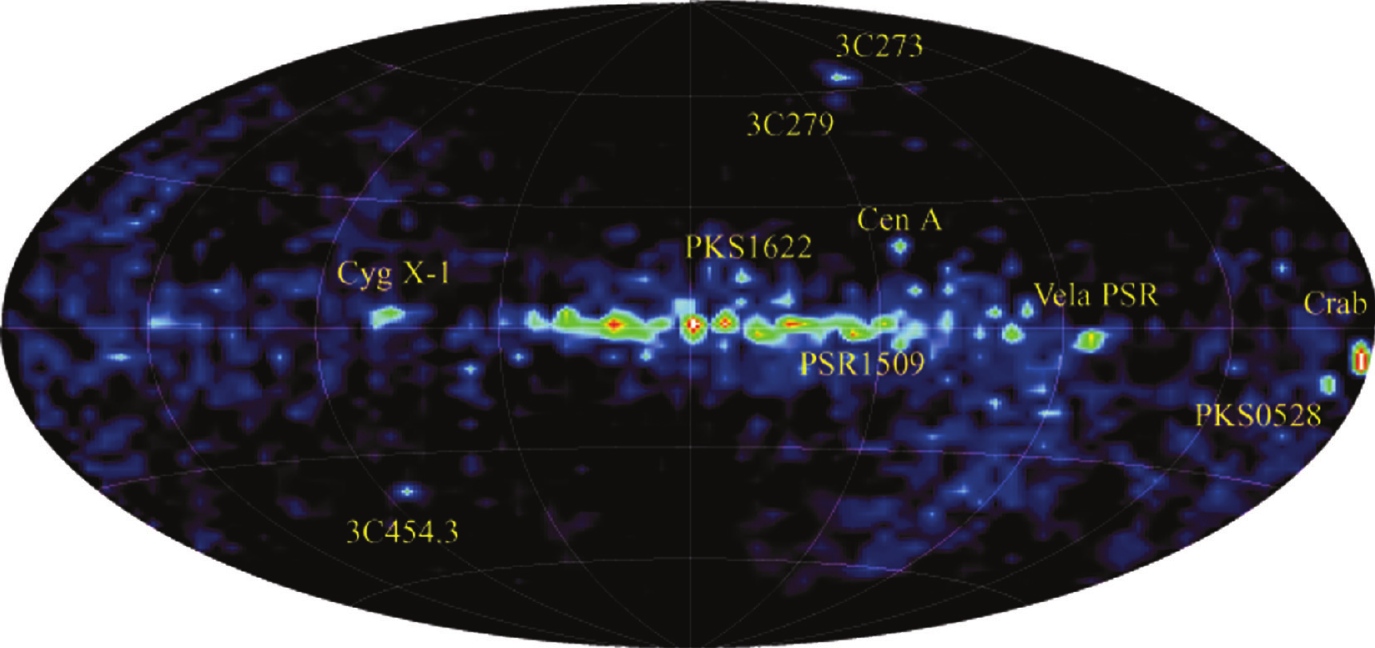}
\caption{The COMPTEL $1-30$ MeV sky map.}
\label{fig:5}  
\end{figure}

The history of GRB research has been
almost entirely     written    by high-energy,
space-borne instruments.
  GRBs were discovered as a phenomenon in
1967   by the {\it Vela} satellites
(Klebesadel, Strong, \& Olson 1973), 
 which were  initially launched
to verify the atmospheric nuclear
test ban treaty.
 For many years GRB research was strongly impeded
   by the fact that
 the distances to the sources were  completely unknown.
   This began to change with the launch of {\it CGRO}
  in 1991 and the use of BATSE and OSSE
  (Kurfess et al. 1992, 1997).
 Within a few months, 
 the uniform sky distribution
   of GRBs made it clear 
 that there was no concentration 
of sources to the Galactic plane, 
and therefore GRBs had to lie 
at either much smaller
     ($\sim$solar system), or
much larger ($\sim$cosmological)
  distances.
  The tradition of outstanding GRB research
  begun by BATSE
  has been continued with the BAT on {\it Swift},
   which is yielding $\sim3-8$ arcmin 
 localizations
   using a shadow mask design 
 with CdZnTe detectors
(Barthelmy et al. 2005).
   One of {\it Swift}'s 
  major successes 
has been the first
  position determination of a short GRB 
(Gehrels et al. 2005).

The {\it INTEGRAL} sky as seen by the IBIS instrument
has built upon the previous COMPTEL results.
  An all-sky map
  was produced which represented
  the first complete scan
    of the entire central
 equatorial zone of the Galactic plane
 at energies above 20 keV  
  with  a limiting sensitivity 
  of $\sim1$ mCrab.
     in the central radian
  (Bird et al. 2006).
 The resultant catalog also included a substantial 
  coverage of extragalactic fields, and
  contained more than 200 high-energy sources.
   The mean position error for sources
   detected with significance above $10\sigma$ was  $\sim40$ arcsec, 
  enough to identify most of them with a known X-ray counterpart
  (Bird et al. 2006).
  This is important given the high Galactic aborption 
  inherent in  most
  earlier surveys.
Lebrun et al. (2004)
   found that the $20-200$ keV
Galactic emission is resolved into compact sources.
  High resolution imaging ($\sim10$ arcmin) of the 
  Galactic center at high energies 
  is a very recent 
  development in the field of high-energy astrophysics. 
 B\'elanger et al. (2006)
  studied the 
    persistent high-energy flux from the 
  Galactic center
   using
  the
 the IBIS/ISGRI imager on {\it INTEGRAL} (Figure 6).
  For the first time they detected a hard X-ray source, 
 IGR J17456-2901, located within $1$ arcmin 
 of  Sagittarius A* (Sgr A*) in 
 the energy range $20-100$ keV.
   By 
  combining the ISGRI spectrum 
  with an X-ray spectrum 
  corresponding to the same physical region around Sgr A* from 
  {\it XMM-Newton} taken during 
 part of the {\it INTEGRAL} observations, 
   they inferred  a  $20-400$ keV 
  luminosity at 8 kpc of  
  $L=(5.37\pm0.21)\times 10^{35}$ erg s$^{-1}$. 
 They also set a  
 $3\sigma$ upper limit on the flux at the 
 electron-positron
 ($e^{-}-e^+$)
   annihilation energy of 
   511 keV from the direction of Sgr A* 
   at $1.9\times10^{-4}$ photons cm$^{-2}$ s$^{-1}$. 
  \begin{figure}
\includegraphics[height=100mm]{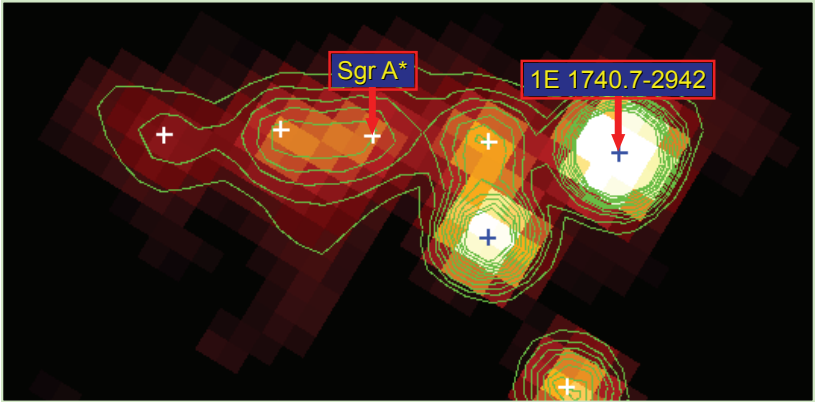}
\caption{View of the Galactic center by {\it INTEGRAL}-IBIS.}
\label{fig:6}  
\end{figure}

\section{Conclusion}

Gamma-ray astronomy  has taken
   enormous strides
  in the last few decades.
Dramatic improvements in both
spectral and imaging capabilities
   have thrust $\gamma-$ray
telescopes into the forefront of 
  astronomical research.
For about a decade 
the prompt localizations
  so essential to
 GRB research were
   carried out largely by
{\it CGRO}; now the torch has been
passed  to {\it Swift}.
  The Galactic center has now been 
 explored in a new and interesting way, via $\gamma-$rays,
  and our knowledge of transient sources
  in the Galactic bulge and
  the diffuse Galactic background has deepened.
   Future missions that seek to take advantage
  of the  ultra-high resolution 
  potential of $\gamma-$ray observing at the
diffraction limit offer the hope of direct imaging of
nearby black hole event horizons - which would allow
us to probe directly General Relativity in the 
strong field limit.




\end{document}